\documentclass[[12pt, letter]{article}

\usepackage{amsmath,amssymb,bm,epsfig,afterpage}
\usepackage{cite}
\usepackage{here}
\usepackage{color}
\usepackage{colortbl}
\usepackage{xcolor}
\usepackage{tabularx}
\usepackage{subcaption}
\usepackage{bbm}
\usepackage{graphicx}
\usepackage{listings}
\usepackage{longtable}
\usepackage[utf8]{inputenc}
\usepackage{cancel}
\usepackage{slashed}

\begin{document}

\begin{titlepage}

\vspace{1.2cm}
\begin{center}
{\Large
{\bf The Absence of Observable Proton Decay in a Global $SU(5)$ F-theory Model
}
}
\vskip 2cm
Herbert Clemens$^{a}$~\footnote{clemens.43@osu.edu}
and
Stuart Raby$^b$~\footnote{raby.1@osu.edu}

\vskip 0.5cm

{\it $^a$
Mathematics Department, Ohio State University, Columbus OH 43210,
USA}\\[3pt]

{\it $^b$
Department of Physics, Ohio State University, Columbus, Ohio, 43210, USA}\\[3pt]

\vskip 1.5cm

\begin{abstract}
We begin with an $E_{8}\times E_{8}$ Heterotic model broken to an
$SU(5)_{gauge} \times U(1)_{X}$ and a twin $SU(5)_{gauge} \times U(1)_{X}$, where one $SU(5)$
and its spectrum is identified as the visible sector while the other
can be identified as a hidden twin sector. In both cases we obtain
the minimal supersymmetric standard model (MSSM) spectrum after Wilson-line
symmetry-breaking enhanced by a low energy R-parity and $\mathbb{Z}_4^R$ symmetry. 
We argue that there will not be any observable proton decay in this model
\end{abstract}

\end{center}

\end{titlepage}

\setcounter{footnote}{0}

\section{Introduction}

Supersymmetric grand unified theories {[}SUSY GUTs{]} \cite{Dimopoulos:1981yj,Dimopoulos:1981zb,Ibanez:1981yh}
have many nice properties. These include an explanation of the family
structure of quarks and leptons with the requisite charge assignments
under the Standard Model {[}SM{]} gauge group $SU(3)_{C}\times SU(2)_{L}\times U(1)_{Y}$
and a prediction of gauge coupling unification at a scale of order
$10^{16}$ GeV. The latter is so far the only direct hint for the
possible observation of supersymmetric particles at the LHC. UV completions
of SUSY GUTs in string theory also provide a consistent quantum mechanical
description of gravity. As a result of this golden confluence, many
groups have searched for SUSY GUTs in string theory. In fact, it has
been shown that by demanding SUSY GUTs in string constructions one
can find many models with features much like that of the minimal supersymmetric
Standard Model {[}MSSM{]} \cite{Lebedev:2006kn,Lebedev:2007hv,Kim:2007mt,Lebedev:2008un,Blumenhagen:2008zz,Anderson:2011ns,Anderson:2012yf}.

The past several years have seen significant attention devoted to
the study of supersymmetric GUTs in $F$-theory \cite{Donagi:2008ca,Beasley:2008dc,Beasley:2008kw,Donagi:2008kj,Blumenhagen:2009yv,Grimm:2010ez,Tatar:2009jk}.
Both local and global $SU(5)$ $F$-theory GUTs have been constructed
where $SU(5)$ is spontaneously broken to the SM via a non-flat hypercharge
flux. One problem with this approach for GUT breaking is that large
threshold corrections are generated at the GUT scale due to the non-vanishing
hypercharge flux \cite{Beasley:2008kw,Donagi:2008kj,Mayrhofer:2013ara,Blumenhagen:2009yv,Blumenhagen:2008aw}.
An alternative approach to breaking the GUT group is using a Wilson
line in the hypercharge direction, i.e. a so-called flat hypercharge
line bundle. In this case it is known that large threshold corrections
are not generated at the GUT scale (or, in fact, it leads to precise
gauge coupling unification at the compactification scale in orbifold
GUTs)\cite{Krippendorf:2013dqa,Raby:2009sf} and \cite{Ross:2004mi,Hebecker:2004ce,Trapletti:2006xv,Anandakrishnan:2012ii}.

\section{The Model}
In this letter we discuss baryon number violation in the model of Refs. \cite{Clemens:2019wgx,Clemens:2019mvs,Clemens:2019dts,Clemens:2019flx,Clemens:2020grq}.
In this F-theory model starting with $E_{8}\times E_{8}$, each $E_8$ is broken to $SU(5)_{gauge} \times U(1)_{X}$ by a 4 + 1 split spectral cover.   This is equivalent to first breaking $E_8$ to $SO(10)$ and then breaking $SO(10)$ to
$SU(5)_{gauge} \times U(1)_{X}$. After a  $\mathbb{Z}_2$ involution which acts freely on the GUT surface, the GUT surface is an Enriques surface with a fundamental group $\Pi_1(S_{GUT}) = \mathbb{Z}_2$.  Simultaneously, $SU(5)$ is broken to the Standard Model gauge group via a hypercharge Wilson line wrapping the GUT surface.   The model has 3 families
of quarks and leptons which reside in the spinor representation of $SO(10)$ and transform as a $10$ and $\bar 5$ representation of $SU(5)$.   Under the involution, $C_{u,v}$, the $10$ and $\bar 5$ representations split into two seperate states which are either even ($10_+, \bar 5_+$) and odd ($10_-, \bar 5_-$) under the involution.   In addition, these states are either even (+) or odd (-) under the Wilson line, $L_Y$. The massless states which remain after the involution transform as either  $(+ +)$ or $(- -)$ under the combined $\mathbb{Z}_2$ involution and Wilson line.\footnote{The massless states which transform as $(+ -)$ and $(- +)$ are projected out of the theory.}  The dimension of the respective cohomologies,  $h^0, h^1$, gives the number of sections in the $10$ and $\bar 5$ representations on the respective matter curves.  See the Tables below.

Note, the quark and lepton doublets are contained in the $(- -)$ sectors, while the $SU(2)$ singlet states are in the $(+ +)$ sector, i.e. the doublets are in a Pati-Salam $(4,2,1)$, while the $SU(2)$ singlets are in a $(\bar 4, 1, 2)$. This means that the $SU(5)$ gauge bosons cannot mediate proton decay at the tree level since the resulting massless states in the $10$ and $\bar 5$ come from different $10$s and $\bar 5$s. Only the Pati-Salam gauge bosons in $SU(4)\times SU(2)_L \times SU(2)_R$ act on these states. But these cannot mediate proton decay at the tree level either. 
Therefore there are no tree level dimension 6 operators mediating proton decay.

What about dimension 4 or 5 baryon and lepton number violating operators? The dimension 4 operators are absent due to R-parity in the model. And the dimension 5 operators are either absent or severely suppressed due the $\mathbb{Z}_4^R$ symmetry \cite{Lee:2010gv}. The bottom line is that proton decay is not observable in this model.
\smallskip{}
\noindent \begin{center}
\begin{tabular}{|c|c|c|c|c|} 
\hline
$\Sigma_{\mathbf{10}}^{\left(4\right)}=\left\{ a_{5}=z=0\right\} $  & $C_{u,v}$  & \textbf{$L_{Y}$} & $\mathcal{L}_{Higgs}$  & $SU\left(3\right)\times SU\left(2\right)\times U\left(1\right)_{Y}$\tabularnewline
\hline
\hline
$h^{0}\left(\check{\mathcal{L}}_{\mathbf{10}}^{\left(4\right)\left[\text{\textpm}1\right]}\right)$ & $+1$  & $+1$  & $3$  & $\left(\mathbf{\mathbf{\mathbf{1}}},\mathbf{1}\right)_{+1}$\tabularnewline
\hline
 & $-1$  & $-1$  &  & $\left(\mathbf{\mathbf{\mathbf{3}}},\mathbf{2}\right)_{+1/6}$\tabularnewline
\hline
 & $+1$  & $+1$  &  & $\left(\mathbf{\mathbf{\mathbf{\bar{3}}}},\mathbf{1}\right)_{-2/3}$\tabularnewline
\hline
$h^{1}\left(\check{\mathcal{L}}_{\mathbf{10}}^{\left(4\right)\left[\text{\textpm}1\right]}\right)$  & $+1$  & $+1$  & $0$  & $\left(\mathbf{\mathbf{\mathbf{1}}},\mathbf{1}\right)_{+1}$\tabularnewline
\hline
 & $-1$  & $-1$  &  & $\left(\mathbf{\mathbf{\mathbf{\bar{3}}}},\mathbf{2}\right)_{+1/6}$\tabularnewline
\hline
 & $+1$  & $+1$  &  & $\left(\mathbf{\mathbf{\mathbf{3}}},\mathbf{1}\right)_{+2/3}$\tabularnewline
\hline
\end{tabular}\smallskip{}
\begin{tabular}{|c|c|c|c|c|} 
\hline
$\Sigma_{\mathbf{\bar{5}}}^{\left(41\right)}=\left\{ a_{420}=z=0\right\} $  & $C_{u,v}$  & \textbf{$L_{Y}$} & $\mathcal{L}_{Higgs}$  & $SU\left(3\right)\times SU\left(2\right)\times U\left(1\right)_{Y}$\tabularnewline
\hline
\hline
$h^{0}\left(\check{\mathcal{L}}_{\mathbf{\bar{5}}}^{\left(41\right)\left[\text{\textpm1}\right]}\right)$  & $+1$  & $+1$  & $3$  & $\left(\mathbf{\mathbf{\mathbf{\bar{3}}}},\mathbf{1}\right)_{+1/3}$\tabularnewline
\hline
 & $-1$  & $-1$  &  & $\left(\mathbf{\mathbf{\mathbf{1}}},\mathbf{2}\right)_{-1/2}$\tabularnewline
\hline
$h^{1}\left(\check{\mathcal{L}}_{\mathbf{\bar{5}}}^{\left(41\right)\left[\text{\textpm1}\right]}\right)$  & $+1$  & $+1$  & $0$  & $\left(\mathbf{\mathbf{\mathbf{3}}},\mathbf{1}\right)_{-1/3}$\tabularnewline
\hline
 & $-1$  & $-1$  &  & $\left(\mathbf{\mathbf{\mathbf{1}}},\mathbf{2}\right)_{+1/2}$\tabularnewline
\hline
\end{tabular}\smallskip{}
\end{center}

\section{Conclusions}
In this brief letter we have argued that the global $SU(5)$ F-theory model presented in \cite{Clemens:2019dts} does not produce any observable proton decay with either dimension 4, 5 or 6 operators.  Dimension 4 operators are forbidden by R-parity, dimension 5 are forbidden by a $\mathbb{Z}_4^R$ symmetry and dimension 6 are forbidden at tree level due
to the Wilson line breaking of $SU(5)$. Apparently only leptogenesis can be used as a mechanism in the early universe to produce the net matter-anti-matter asymmetry.

\section*{Acknowledgements}
SR thanks Junichiro Kawamura for useful comments and also acknowledges partial
support from Department of Energy grant DE-SC0011726.

\clearpage
{\small
\bibliographystyle{jhep}
\bibliography{pdecay}
}

\end{document}